\documentclass[fleqn,usenatbib]{mnras}

\usepackage{newtxtext,newtxmath}
\usepackage[T1]{fontenc}

\DeclareRobustCommand{\VAN}[3]{#2}
\let\VANthebibliography\thebibliography
\def\thebibliography{\DeclareRobustCommand{\VAN}[3]{##3}\VANthebibliography}

\usepackage{graphicx}	
\usepackage{amsmath}	
\usepackage{siunitx}

\title[Difference in B and Be star binarity]{Gaia uncovers difference in B and Be star binarity at small scales: evidence for mass transfer causing the Be phenomenon}

\author[J. M Dodd et al.]{Jonathan M. Dodd,$^{1}$\thanks{E-mail: py17jd@leeds.ac.uk}
Ren\'{e} D. Oudmaijer,$^{1}$
Isaac C. Radley,$^{1}$
Miguel Vioque,$^{2,3}$
and Abigail J. Frost$^{4,5}$
\\
$^{1}$School of Physics \& Astronomy, University of Leeds, Woodhouse Lane, Leeds, LS2 9JT, UK\\
$^{2}$Joint ALMA Observatory, Alonso de C\'{o}rdova 3107, Vitacura, Santiago, 763-0355, Chile\\
$^{3}$National Radio Astronomy Observatory, 520 Edgemont Road, Charlottesville, VA 22903, USA\\
$^{4}$Institute of Astronomy, KU Leuven, Celestijnlaan 200D, Leuven, 3001, Belgium\\
$^{5}$European Southern Observatory (ESO), Alonso de C\'{o}rdova 3107, Vitacura, Santiago, 763-0355, Chile
}

\date{Accepted XXX. Received YYY; in original form ZZZ}

\pubyear{2023}

\begin{document}

    \label{firstpage}
    \pagerange{\pageref{firstpage}--\pageref{lastpage}}
    \maketitle

    \begin{abstract}
        Be stars make up almost 20\% of the B star population, and are rapidly rotating stars surrounded by a disc; however the origin of this rotation remains unclear. Mass transfer within close binaries provides the leading hypothesis, with previous detections of stripped companions to Be stars supporting this. Here, we exploit the exquisite astrometric precision of Gaia to carry out the largest to date comparative study into the binarity of matched samples of nearby B and Be stars from the Bright Star Catalogue. By utilising new ``proper motion anomaly’’ values, derived from Gaia DR2 and DR3 astrometric data alongside previous values calculated using Hipparcos and Gaia data, and the Gaia provided RUWE, we demonstrate that we can identify unresolved binaries down to separations of 0.02$\si{''}$. Using these measures, we find that the binary fractions of B and Be stars are similar between 0.04 – 10$\si{''}$, but the Be binary fraction is significantly lower than that of the B stars for separations below 0.04$\si{''}$. As the separation range of these ``missing’’ binaries is too large for mass transfer, and stripped companions are not retrieved by these measures, we suggest the companions migrate inwards via binary hardening within a triple system. This confirms statistically for the first time the hypothesis that binary interaction causes the Be phenomenon, with migration causing the dearth of Be binaries between 0.02 - 0.04$\si{''}$. Furthermore, we suggest that triplicity plays a vital role in this migration, and thus in the formation of Be stars as a whole.
    \end{abstract}


    \section {Introduction} \label{sec:introduction}

        Be stars make up around $20\%$ of the total population of B-type stars \citep{J_Bodensteiner_2020} and are defined, in part, by the presence of Balmer emission lines. These lines are thought to originate from a slowly outflowing ionised circumstellar gas-disc \citep{J_Porter_2003, T_Rivinius_2013, C_E_Jones_2022}. While the dynamics of these discs have been studied extensively \citep{J_P_Wisniewski_2010, Z_H_Draper_2014, M_Suffak_2022}, the mechanism by which they form is undetermined. Consensus suggests that the formation is related to the rapid rotation observed in practically the entire Be population \citep{J_Zorec_2016, J_Zorec_2017}, and which is suggested to be at or near the critical velocities of such stars \citep{R_H_D_Townsend_2004, A_Granada_2013}. Such rapid rotation could thus allow for processes such as turbulence \citep{R_H_D_Townsend_2004} or non-radial pulsations \citep{S_P_Owocki_2002,T_Semaan_2018} to contribute significantly to the formation of the decretion discs around Be stars due to decreased effective gravity at the surface of such stars \citep{A_Granada_2013}. This, however, only leads to a further question - what exactly causes Be stars to rapidly rotate?
        
        Three mechanisms have been suggested as the origin of this characteristic rotation; it could be a relic from the parent molecular cloud, the rotation thus a result of inherited angular momentum \citep{P_Bodenheimer_1995}; the envelope of the star could be spun up via momentum transport from a contracting core \citep{A_Granada_2013, B_Hastings_2020}; or, crucially for this study, it could be a result of mass and angular momentum transfer via binary interaction \citep{Y_Shao_2014, J_Bodensteiner_2020}.

        If binary interaction is the source of this rapid rotation, then one could expect that the binary statistics of B and Be stars will differ. This line of reasoning provided the basis for the study performed in \cite{R_Oudmaijer_2010}, which found the binary fraction of B and Be-type stars to be similar between separations of $0.1$ and $8\si{''}$ based on a sample of $36$ B and $37$ Be stars, and concluded that binarity, at least at these scales, could not be responsible for all Be stars. However, while a number of studies into the binarity of Be stars exist (\citealt{D_Boubert_2018, R_Klement_2019, B_Hastings_2021} and \citealt{K_ElBadry_2022}, among others) and suggest the stripping of the companion to the Be star as the cause \citep{Z_Han_2002, Y_Naze_2022, K_ElBadry_2021b, K_ElBadry_2022}, very few comparative studies into the binarity of both B and Be-type stars exist. The next most recent points of comparison are \cite{H_A_Abt_1978} and \cite{H_A_Abt_1984}, each from around 40 years ago, and which similarly conclude that binarity cannot be responsible for the Be phenomenon at the scales they probe. However, \cite{H_A_Abt_1978}, which utilises radial velocity variations to detect binarity, suffers from the effects of small sample sizes, while \cite{H_A_Abt_1984} is hampered by inhomogeneous biases as it is, in effect, a review of the then available literature.

        With the advent of high-precision astrometric all-sky surveys, we are now able to mitigate these issues and investigate the multiplicity of large samples. Recent studies have employed astrometric data from surveys such as Hipparcos \citep{F_vanLeeuwen_2007} and Gaia \citep{Gaia_Collaboration_2021} in order to detect binarity in the overall stellar population, both in the case of wide \citep{K_El_Badry_2021} and close \citep{P_Kervella_2019a, P_Kervella_2021} binaries. The time is thus ripe for such an investigation into the binarity of B and Be stars.

        Here we utilise the Proper Motion Anomaly (PMa), as introduced in \cite{P_Kervella_2019a} (see also \citealt{T_Brandt_2018}), which is a measure of the difference between long and short term proper motions of a point source, in order to determine the fractions of B and Be type stars in visually unresolved binary systems. Alongside previously determined PMa values from \cite{P_Kervella_2021}, calculated using Hipparcos and Gaia DR3 data (a similar catalogue can also be found in \citealt{T_Brandt_2021}), we use new values, calculated for the first time in this work using a combination of Gaia DR2 and DR3 data \citep{Gaia_Collaboration_2018, Gaia_Collaboration_2022}. Using these metrics we probe binary systems with separations between $0.02$ and $1.1\si{''}$, and with magnitude differences up to $4\si{mag}$.
        
        Additionally we utilise the Gaia-provided Renormalised Unit Weight Error (RUWE) to further constrain the binary fractions of these two populations \citep{K_Stassun_2021}. Thus we are able to bridge the gap in binary separations between, and expand upon previous radial velocity \citep{H_A_Abt_1978} and adaptive optics \citep{R_Oudmaijer_2010} studies of these stars. Additionally, by querying a sample over an order of magnitude larger than present in the most recent similar study \citep{R_Oudmaijer_2010}, this paper provides the largest scale comparative study of B and Be star binarity to date.
        
        This paper is structured as follows: in Section \ref{sec:methods} we detail the two binary detection methods used (the PMa and the RUWE), and demonstrate their sensitivities and limitations; in Section \ref{sec:results} we report the results of utilising these methods on samples of B and Be stars; and in Section \ref{sec:discussion} we discuss the implications of these newly reported binary fractions on models for the formation of Be stars.
    
    \section{Methods}\label{sec:methods}

        \subsection{Binary Detections via Proper Motion Anomaly} \label{section:pma}

            For any binary system we can define two similar variables, the centre of mass, which the two component stars orbit, and the photocentre, which describes the location of the apparent point source in an unresolved system. In general, the positions of these two points differ, thus the photocentre also orbits the centre of mass - leading to a deviation from the linear proper motion expected for single sources in the case of unresolved binary systems.

            \begin{figure}
                \centering
                \includegraphics[width = 0.4\textwidth]{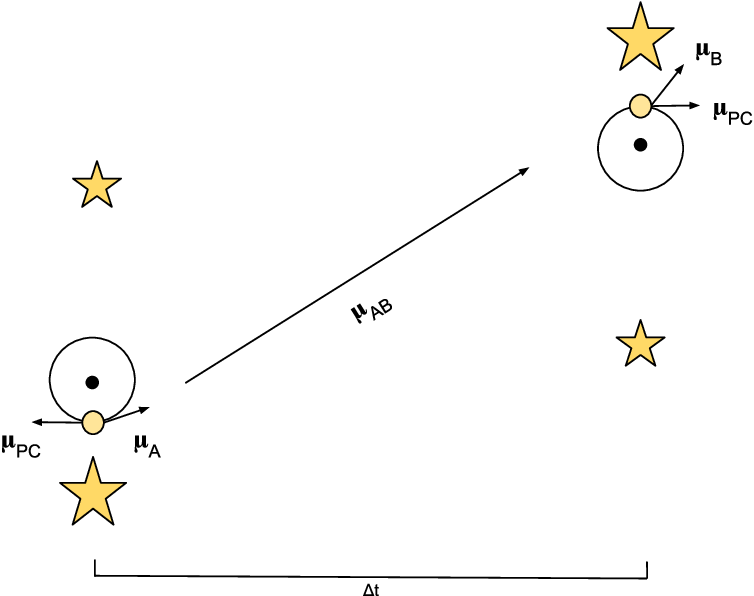}
                \caption{Cartoon illustrating the principle behind the Proper Motion Anomaly. The large and small stars are the primary and secondary companion respectively, the black dot is the centre of mass and the yellow dot the photocentre. Two sets of observations are taken at $A$ and $B$, with a time interval of $\Delta t$ between them. $\mu_A$ and $\mu_B$ are the short term proper motions measured during their respective observations, and $\mu_{AB}$ is the long term proper motion, determined from the change of observed position over $\Delta t$. Arrows labelled $\mu_{PC}$ are the instantaneous proper motions of the photocentre in the frame which the centre of mass is stationary.}
                \label{fig:pma_diagram}
            \end{figure}
        
            This deviation from the expected linear proper motion, referred to as the Proper Motion Anomaly (PMa), $\Delta\mu$, can be detected through a comparison of long and short term proper motion measurements ($\mu_{lt}$ and $\mu_{st}$ respectively), with the long term effectively tracing the motion of the centre of mass and the short term tracing the sum of the centre of mass and relative photocentre motions. An illustration of this principle can be found in Figure \ref{fig:pma_diagram}. The PMa can thus be expressed as:

            \begin{equation}
                \label{eq:pma}
                \Delta\mu = \mu_{st} - \mu_{lt}
            \end{equation}
            
            Thanks to current advances in the precision of proper motion measurements, alongside high precision positional measurements allowing for long term proper motions to be calculated, the PMa thus provides a powerful tool in the search for unresolved binary candidates within large samples of stars.

            This measure is sensitive to the period of a given binary system. Periods significantly longer than the time elapsed over the long term proper motion measurement, $\Delta t$ will have a deviation in proper motion too small to be detected. Similarly, systems wherein a whole number of complete orbits have occurred during the measurement of the long term proper motion will not be detected, again as the deviation in long and short term proper motions will be too small to detect.

            Following this logic, peak sensitivity of the PMa should be achieved for systems where an $\frac{2n + 1}{2}$ orbital periods have elapsed between the two observations, as the absolute change in orbital phase is at a maximum here. However, in reality, due to the time-window smearing effects present over the observation periods of the two constituent astrometric surveys \citep{P_Kervella_2021} there is a general decrease in sensitivity towards smaller periods, with peak sensitivity being achieved for periods of $P \simeq \Delta t$ \citep{P_Kervella_2021}. This also explains why, although of much higher astrometric precision, the DR2-DR3 derived PMa's probe larger periods than the Hipparcos-DR3 PMa values (as will be shown in the following sub-subsection). Below we will probe the sensitivity limits of the various methods both theoretically and empirically.
        
            \subsubsection{Separation Limits of the Proper Motion Anomaly}\label{sec:pma_sep_lim}

                Whereas the sensitivity of the PMa to mass as a function of orbital period (and thus linear separation) has previously been investigated \citep{P_Kervella_2021}, the sensitivity to angular separation remains unexamined. 
                As detailed in Appendix \ref{sec:derivation}, using an idealised theoretical model, consisting of a face-on, circularly orbiting binary system, we arrive at a minimum detectable separation of $0.01\si{''}$ for PMa values calculated using Hipparcos and Gaia DR3 data. Additionally, when considering the PMa calculated using Gaia DR2 and DR3 data, the theoretical minimum detectable separation is $0.0025\si{''}$.  However, this computation does not take into account the effect of the time-window smearing during the PM measurement - which leads to the DR2-DR3 PMa losing out on the shorter separations.   
                
                It is thus useful to obtain an empirical value as a point of comparison. In order to evaluate this sensitivity of the PMa to varying separations, we use a sample of binary systems listed in the Washington Double Star (WDS, \citealt{WDS}) catalogue. The WDS contains a compilation of multiple star systems for which at least one measured separation is available within the literature.
                The sources in the WDS were cross-matched with the list provided in \cite{P_Kervella_2021} using the xMatch service provided by CDS, Strasbourg \citep{F_Pineau_2020}. No restrictions on mass or spectral type were placed on the stars within the binaries in order to generate the largest, general sample. Reported higher order multiple systems were then removed for the sake of simplicity, as were any systems with separations greater than $10\si{''}$. While the PMa primarily probes the unresolved regime of separations by nature, this $10\si{''}$ limit was selected in order to ensure that no unpredictable effects were materialising at higher separations.

                For the resulting $11\,116$ unique binary systems, we obtained values of the PMa from the \cite{P_Kervella_2021} catalogue (which uses data reported in the Hipparcos re-reduction \citep{F_vanLeeuwen_2007} and the Gaia DR3 catalogues, giving $\Delta t \approx 25\,\si{yrs}$) and calculated new values, using data available in the Gaia Data Release 2 and Data Release 3 (\citealt{Gaia_Collaboration_2018, Gaia_Collaboration_2021}; $\Delta t \approx 0.5\,\si{yrs}$). The methodology used in \cite{P_Kervella_2021} was utilised for the calculation of these new values, with calibration performed via re-calculating Hipparcos-Gaia DR3 PMa's replicating the reported PMa values exactly and reported errors to within a precision of $1\%$.
                
                \begin{figure}
                    \centering
                    \includegraphics[width = 0.45\textwidth]{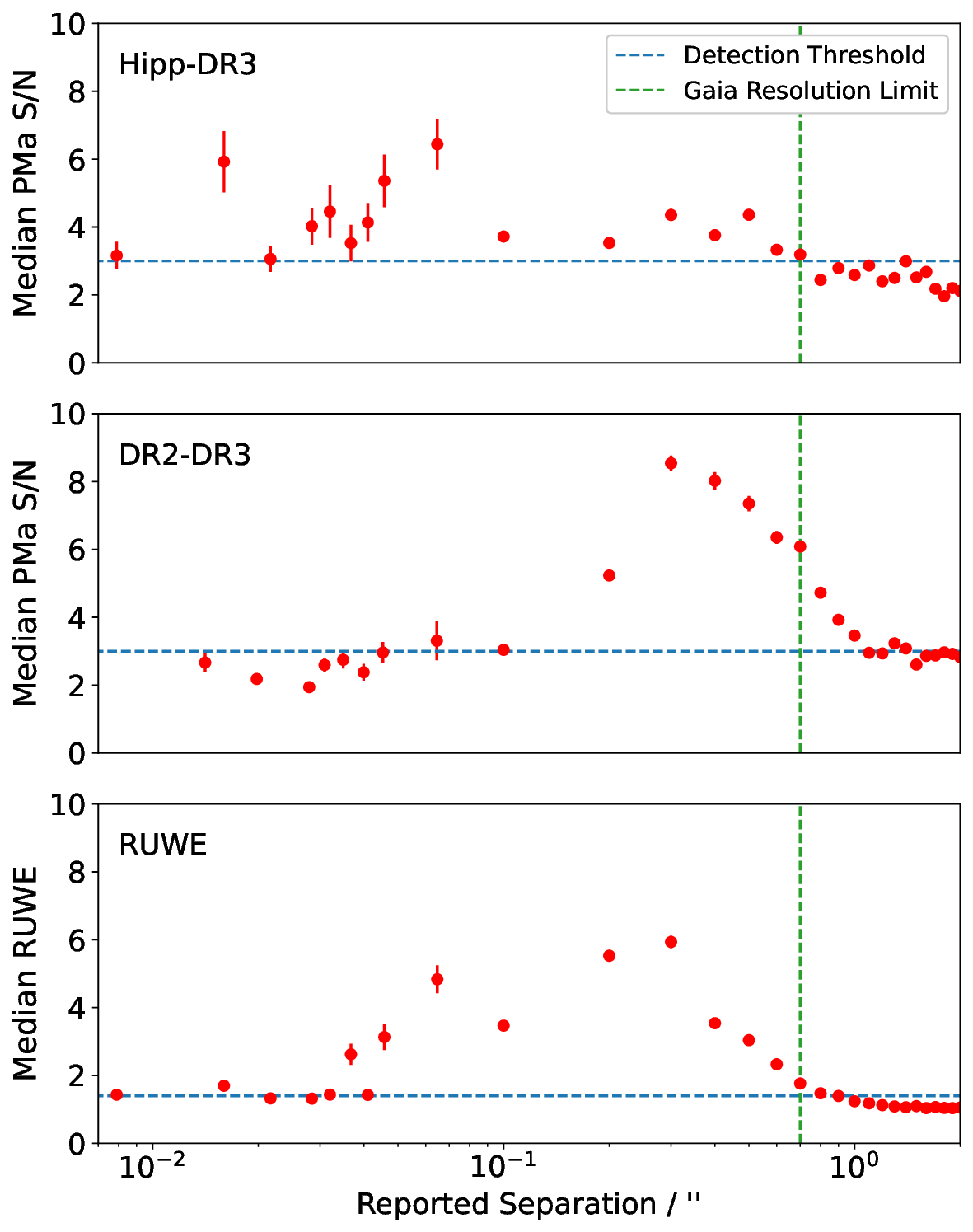}
                    \caption{Median values of the Hipparcos - Gaia DR3 and Gaia DR2 - DR3 PMa S/N's and the RUWE, of systems reported as binary in the WDS \protect\citep{WDS}. Here values greater than 3 constitute a detection of binarity by either PMa, and values greater than 1.4 constitute a detection by the RUWE - see Section \ref{section:ruwe} for details.}
                    \label{fig:sep_lims}
                \end{figure}
           
                The final sample contains $10\,283$ sources with reported PMa values; note that not all values surpass the binary detection threshold of S/N$>$3. The majority of these systems had separations reported to a precision of $0.1\si{''}$. $316$ systems have separations reported as $0\si{''}$ in the WDS Catalogue (i.e. their separations were below $0.1\si{''}$). For each of these systems, higher precision separations were retrieved from supplemental WDS material \citep{WDS}. Where orbital solutions had previously been reported, the separation available at the Gaia DR3 epoch was used.

                Figure \ref{fig:sep_lims} shows the median values of the signal-to-noise ratio for the Hipparcos-Gaia DR3 PMa for the WDS sample at varying separations. These systems have been placed into bins of width $0.1\si{''}$ at separations greater than or equal to $0.1\si{''}$. At separations below $0.1\si{''}$ the systems have been placed into 10 bins such that each bin has an equal number of systems within it (approximately 30 per bin). For these bins the mean reported separation of the systems within it is plotted. 
            
                Based on the data presented in Figure \ref{fig:sep_lims}, it can be seen that for the DR3-Hipparcos PMa, there is an increase of SNR towards lower reported separations. On average, this PMa remains above the detection threshold of $3$ for all separations between $0.02$ and $0.7\si{''}$ (based on the median values). As a result, $0.02 - 0.7\si{''}$ appears to be an empirical range of separations detectable via this PMa;  at least half of all WDS binary systems with separations within this range are detected (note that we wouldn't expect for all binaries within this range to be detected due to projection effects). This is in agreement  with the calculated theoretical minimum detectable separation and the expectation that the upper limit occurs somewhere around the spatial resolution limit of the surveys.

                It is also to be noted that high PMa systems at separations greater than the resolvable limit (i.e. with separations greater than $\sim1\si{''}$) exist in non-negligible numbers. These systems have reported separations much greater than the PMa can detect (as it measures the motion of an essentially unresolved binary), and it is thus likely that they are, in fact, as of yet undetected higher order multiple systems.
                
                Alongside the examination of the \cite{P_Kervella_2021} reported Hipparcos-Gaia DR3 PMa values, an equivalent consideration of Gaia DR2-DR3 PMa values has been conducted. As there does not exist an equivalent catalogue as in \cite{P_Kervella_2021}, these PMa's were thus calculated via the same process, but using data from the Gaia Data Releases 2 and 3.
        
                The median values of these PMa SNR values at a given separation can also be found in Figure \ref{fig:sep_lims}. Again, a rise in the median PMa can  be seen towards lower separations, here beginning at separations of approximately $1.1\si{''}$, increasing towards a peak at around $0.6\si{''}$, and remaining above the detection threshold down to $0.2\si{''}$. While this lower limit is in disagreement with the theoretical limit of $0.0025\si{''}$ derived in Appendix \ref{sec:derivation}, the theoretical considerations largely ignore the effect of smearing over the observing timeframe. As the DR2-DR3 PMa is smeared across the entirety of the long-term proper motions timeframe, this will mean that the lowest separations are not probed, although the number of potential observable objects increases significantly (as Gaia has observed orders or magnitude more objects than Hipparcos). Additionally the increase in precision achieved since Hipparcos has both resulted in the upper limit of separations detectable to be expanded beyond the Gaia resolution limit, and increased the total fraction of systems detected within the sensitive limit.

                These results remain consistent when examining B and Be type binaries within the WDS \citep{WDS}, with systems within the two sensitive ranges being detected at significantly higher rates than those outside that range.

                We thus show that the Hipparcos-Gaia DR3 PMa is sensitive to binary systems with angular separations between $0.02 - 0.7\si{''}$, and the Gaia DR2 - DR3 PMa is sensitive between separations of $0.2 - 1.1\si{''}$.                
                
            \subsubsection {Magnitude Limits of the PMa}
        
                \begin{figure}
                    \centering
                    \includegraphics[width = 0.45\textwidth]{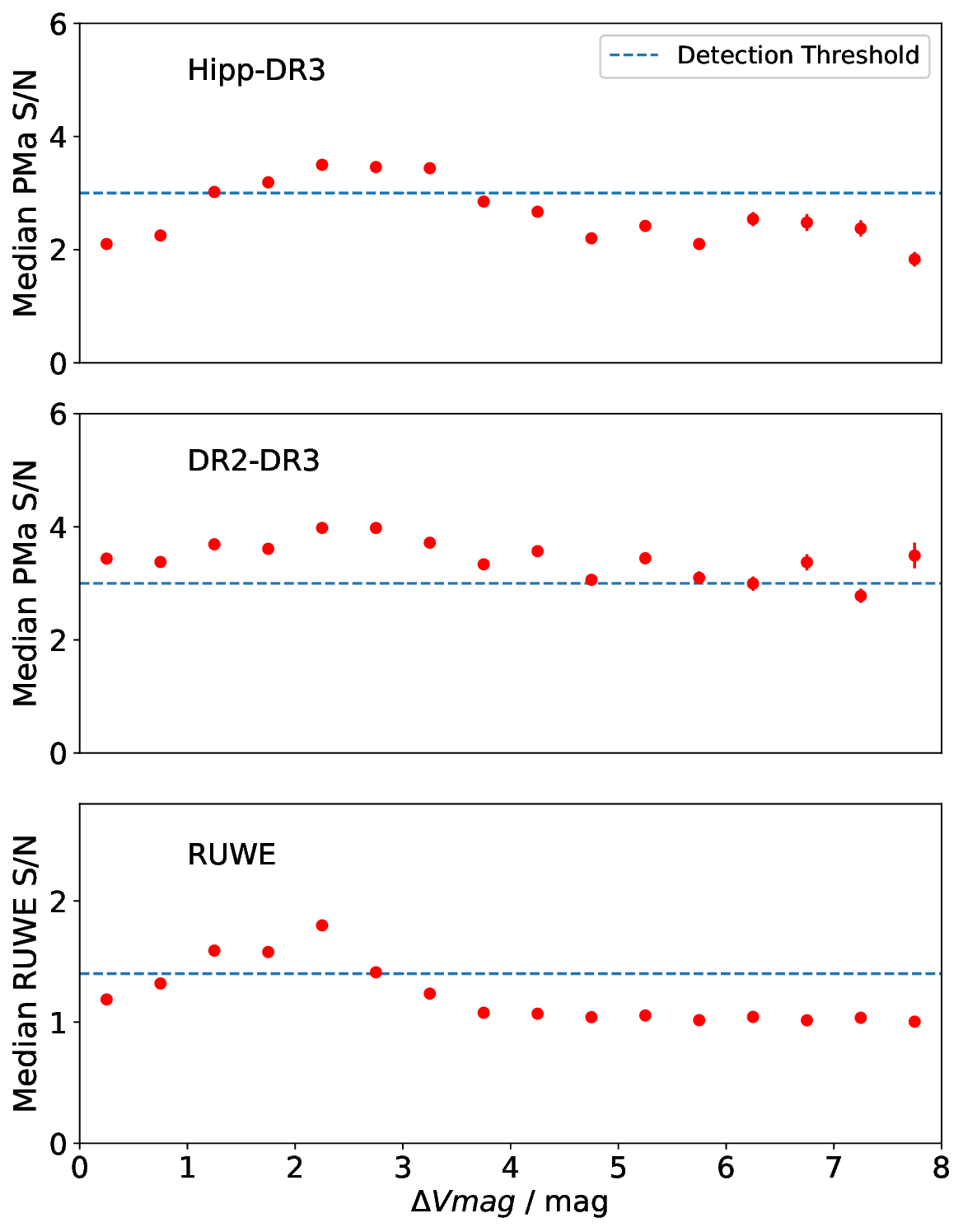}
                    \caption{Median PMa of WDS systems, as a function of the difference in magnitude of the two stars for the Hipparcos-Gaia DR3 and Gaia DR2-DR3 PMa's and the RUWE. Both plots have been cropped to a maximum difference in magnitude of $8\si{mag}$, as the number of systems with magnitude differences greater than this decreases to the point where meaningful conclusions cannot be drawn. Detections are reported out to differences of up to $12\si{mag}$ in all three cases.}
                    \label{fig:mag_lim}
                \end{figure}
            
                In order to evaluate the limitations of the Proper Motion Anomaly, we must also consider the sensitivity of this measure to the difference in magnitude between the two components. Here systems contained within the WDS catalogue are once again considered, now limiting the sample to those systems with reported separations between $0$ and $1\si{''}$ in line with the more favourably sensitive range determined in the previous section.
            
                In the case of the Hipparcos-Gaia DR3 PMa, and as can be seen in Figure \ref{fig:mag_lim}, examining the difference in the WDS reported apparent magnitude reveals that the PMa shows some preferential sensitivity within the $1.0$ to $3.5\si{mag}$ range, peaking at around $\Delta V_{mag} = 3\si{mag}$.
                
                In the case of the Gaia DR2-DR3 PMa, an equivalent examination reveals a general increase in sensitivity across most magnitude differences available via the WDS sample, with the preferentially detected region expanding to systems with a difference in magnitude $0 < \Delta V_{mag} \leq 4.5$, peaking at around $\Delta V_{mag} = 2.5$, although significant numbers of detections are still made at greater differences in magnitude.

        \subsection{Binary Detections via the RUWE} \label{section:ruwe}

            The Renormalised Unit Weight Error, or the RUWE, is a reduced chi-squared parameter that describes the `goodness-of-fit' of a single body astrometric model to an observed source. As such, it would follow that such a parameter is sensitive to the presence of unresolved binary companions - with the photocentre wobble caused by such a companion causing deviations from what is expected of a single body, resulting in a worse fit astrometric solution. Indeed, the RUWE parameter has been suggested to act as a binarity indicator \citep{V_Belokurov_2020}, with values significantly above the normalised value of $1$ (i.e $\geq1.4$ as per \citealt{K_Stassun_2021}) being regarded as indicative of the source being `badly fit' by the astrometric solution and thus potentially binary. 
        
            Like the PMa, the RUWE varies in sensitivity according to the period of a binary system, with the RUWE having some function of sensitivity to binarity determined both by the period of the system in question and by the time baseline over which the RUWE is determined (approximately six months in the case of Gaia DR2 and DR3), with \cite{Z_Penoyre_2021} suggesting that for Gaia DR3, RUWE is an effective measure of binarity for periods of around a month to a decade.

            However, due to sensitivity to other (non-multiplicity related) factors which can inflate the RUWE \citep{V_Belokurov_2020}, it is important to note that there may be some fraction of false positive detections - for example if the sample contains stars known to have circumstellar discs or extended regions of emission \citep{R_Oudmaijer_2022}. This is unlikely to hamper our study of B and Be stars however, as the discs of Be stars are very small when compared to the resolution achieved by Gaia DR3 \citep{A_Quirrenbach_1997, H_Wheelwright_2012}.
        
        \subsubsection {Separation Limits of the RUWE} \label{sec:ruwesep}
        
            The separation limits of the RUWE were determined using the same process as for the PMa above. Here, the detection threshold is given to be $1.4$ (as suggested by \citealt{K_Stassun_2021}).
            Based on the data presented in Figure \ref{fig:sep_lims}, we find that the RUWE exhibits a sharp and marked increase above the detection threshold at binary separations below the resolution limit of Gaia ($0.7\si{''}$). This increase peaks at approximately $0.09\si{''}$, and, on average, remains above the threshold down to separations of $0.04\si{''}$. Thus, we find that the RUWE is sensitive to those binary systems with separations ranging between $0.04\si{''}$ and $0.7\si{''}$.
  
        \subsubsection {Magnitude Limits of the RUWE}
        
            As with the PMa, it is useful to quantify the sensitivity of the RUWE to the magnitude differences of the binary systems probed. Figure \ref{fig:mag_lim} demonstrates, by repeating the same process as used to determine the magnitude differences the PMa is sensitive to, that it would seem that the RUWE is primarily sensitive to systems with magnitude differences between $1.5$ and $3.0\si{mag}$, peaking somewhere between $2.0$ and $2.5\si{mag}$.

            We have thus shown that, via these astrometric methods, we can consistently detect visually unresolved binaries with separations as low as $0.04\si{''}$ using the RUWE, and $0.02\si{''}$ using the Hipparcos-Gaia PMa and with magnitude differences up to $4.5\si{mag}$, depending on the measure used. Thus, thanks to the precision achieved by modern astrometric surveys, we will be able to probe the origins of Be stars with a statistical basis heretofore unavailable.
    
    \section{The Binarity of B and Be Stars}\label{sec:results}

        \subsection{Sample Selection} \label{sec:sample}

            Here we utilise the same initial sample of B and Be stars extracted from the Bright Star Catalogue (BSC) 5th Revised Edition \citep{BSC, BSC_Supplement} as used in Radley et al. (in preparation). This sample was then cross-matched within $5\si{''}$ against the Gaia Data Release 3 \citep{Gaia_Collaboration_2021} data. Stars without proper motion measurements in the Hipparcos or Gaia Data Releases were removed. Stars with non-V luminosity classes were also removed in order to ensure that the comparisons made between our two samples are between like sources that are subject to the same selection effects. A summary of the spectral types of the stars included in this sample can be found in Table \ref{tab:sample}.

            \begin{table}
                \centering
                \caption{Populations of main sequence B and Be type stars within the sample, divided by spectral type.}
                \begin{tabular}{c|c|c}
                    \textbf{Spectral Type} & \textbf{B} & \textbf{Be} \\ \hline 
                                   B0V     &         12 & 4           \\
                                   B1V     &         47 & 10          \\
                                   B2V     &        104 & 31          \\
                                   B3V     &         97 & 19          \\
                                   B4V     &         31 & 12          \\
                                   B5V     &        111 & 13          \\
                                   B6V     &         43 & 8           \\
                                   B7V     &         48 & 4           \\
                                   B8V     &        156 & 13          \\
                                   B9V     &        258 & 9           \\
                                   Total   &        907 & 123         \\
                \end{tabular}
                \label{tab:sample}
            \end{table}
            
            This sample provides an almost thirty-fold increase in the number of B-type stars, and a four-fold increase in the number of Be-type stars compared to similar comparative studies \citep{H_A_Abt_1978, R_Oudmaijer_2010} that probe a well defined separation range. Additionally, by virtue of being selected from the same catalogue of the brightest stars in the sky, these stars will be subject to the same selection effects, and have similar brightnesses and parallaxes, thus ensuring their usefulness in comparing how B and Be type stars differ in their astrometric properties. These stars are also amongst the most nearby ones, thus allowing us to detect the physically closest companions whilst remaining within the range of angular separations detectable by our methods (see Sections \ref{sec:pma_sep_lim} and \ref{sec:ruwesep}), with the lower end of our detectable range, $0.02''$ equating to $5\,\si{au}$ at the mean distance to the stars in these samples.

        \subsection {Detection of Unresolved Binaries}

	        We first consider the complete, unmodified samples of $907$ B and $123$ Be stars, utilising a combination of PMa values found within the catalogue contained within \cite{P_Kervella_2021}, calculated using Hipparcos and Gaia DR3 data, and new PMa values calculated using data reported in Gaia Data Releases 2 and 3, as well as the Gaia-provided RUWE. These measures allow us to detect unresolved binary systems via differences in long and short term proper motions (for the PMa; see Section \ref{section:pma}) and via deviations from a single body astrometric fit (in the case of the RUWE; see Section \ref{section:ruwe}).
         
            Using the Hipparcos-Gaia DR3 PMa, which probes the range 0.02 $-$ 0.7'' (see Section \ref{sec:pma_sep_lim}), equating to around $5\,-\,180\,\si{au} $ at the average distance of $260\si{pc}$, we arrive at binary fractions of $42\pm2\%$ and $28\pm4\%$ for the B and Be samples respectively\footnote{Uncertainties here are calculated using the binomial method. Error bars are given at $1\sigma$.}. This proves to be a greater than $3\sigma$ difference.
            
            In contrast, the Gaia DR2 - DR3 PMa values reveal similar B and Be binary fractions in the 0.2 $-$ 1.1'' range (around $50\,-\,290\,\si{au}$ at $260\si{pc}$). In this case, the fraction of B stars detected as being in a binary system is found to be $27\pm1\%$ and the equivalent for Be stars is found to be $29\pm4\%$.

            Finally, the fractions of B and Be stars detected via applying a detection threshold of $RUWE > 1.4$, tracing binaries with separations between 0.04 $-$ 1.1'' (10 $-$ 290$\,\si{au}$), were afound to be $27\pm1\%$ and $20\pm4\%$ respectively. See Table \ref{tab:separation_binary_frac} further down for a summary of these binary fractions.
            
	    \subsection {Correcting Biases} \label{sec:biases}    
        
            Although the global properties of the B and Be stars in the BSC are very similar, the two samples exhibit slightly different properties across a number of observables. For example, by virtue of being rarer, the Be stars are on average further away than the B stars, with the average B star in our sample sitting at a parallax of $5.26\si{mas}$ with a standard deviation of $3.38\si{mas}$, whereas the average Be star sits at a parallax of $3.81\si{mas}$ with a standard deviation of $2.42\si{mas}$. Likewise the average apparent magnitudes of the B and Be populations are $6.06\si{mag}$ (standard deviation of $1.07\si{mag}$) and $5.95\si{mag}$ (standard deviation of $1.32\si{mag}$) respectively. In addition to this, the two populations differ in the distribution of their spectral types (for example, almost 30\% of the B stars are B9V, while only 7\% of the Be stars are B9Ve). As such, the variations in the distributions of these properties must be taken into account to ensure that the binary fractions are not affected by them.
            
            Thus, in order to remove the effects of these biases, a variety of sub-samples of B stars were created. These included random samples, alongside ``semi-random'' samples wherein the B population was binned according to a given variable, mentioned above, and stars were randomly picked from each bin in order to match the distribution of said variable in the Be sample. In each case a Kolmogorov-Smirnov test was performed in order to confirm that the sub-sample was likely drawn from the same distribution as the Be population. The binary fraction of each sub-sample was then determined using the PMa and RUWE detection thresholds detailed above.

\begin{figure*}
                \centering
                \includegraphics[width = 0.9\textwidth]{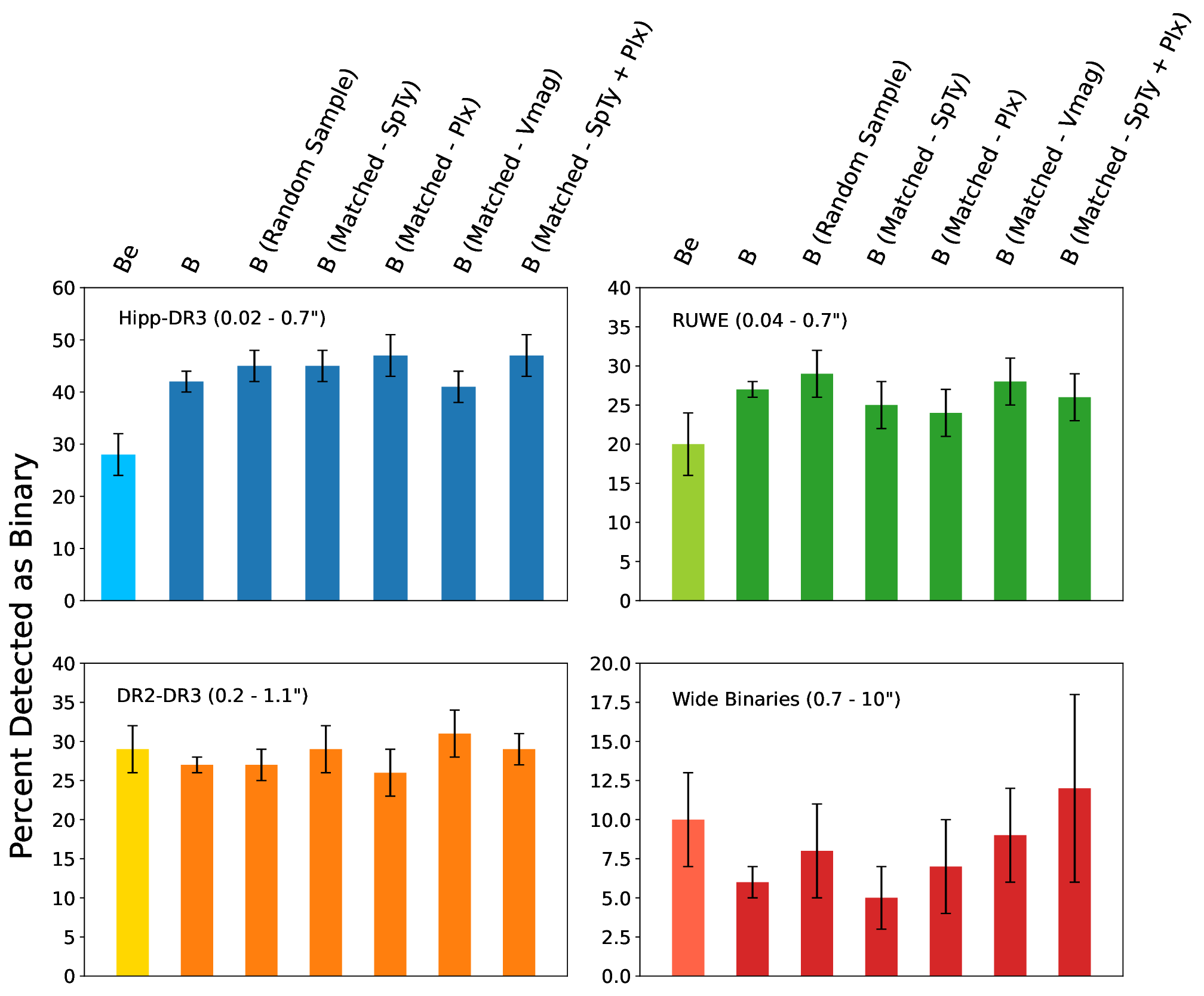}
                \caption{Summary of the binary fractions of B and Be stars, as determined by Hipparcos-Gaia DR3, Gaia DR2 - DR3 PMa values, Gaia DR3 RUWE values and resolved binaries (see text). Also presented are the binary fractions of the various sub-samples of the B type stars, taken in order to account for any biasing effects. `Matched' here indicates that a sub-sample of the B stars has been created such that the distribution of a given variable in the sub-sample is drawn from the same parent distribution as the Be population. Note that the `Wide Binaries' plot is noisier due to smaller number statistics. The Hipparcos - Gaia DR3 values, probing the smallest separations, show a significantly lower Be binary frequency (in each case a greater than $3\sigma$ difference), while all other approaches return similar binary fractions to within the errors. Based upon this we report a dearth of Be binaries with separations between $0.02$ and $0.7\si{''}$, and, as discussed in Section \ref{sec:constraints} we further constrain these missing binaries to between $0.02$ and $0.2\si{''}$.}
                \label{fig:results_summary}
            \end{figure*}

            A summary of these samples and their binary fractions can be found in Figure \ref{fig:results_summary}. Therein, the greater than $3\sigma$ dearth of Be binaries, when compared to the number of B binaries, is still seen to be present at the tightest separations probed by our methods.

        \subsection{Wider Separation Binaries} \label{sec:wide}
        
            While the focus of this study is on close binaries, we now consider the presence of binary companions outside the $0.02 - 0.7\si{''}$ range. \cite{R_Oudmaijer_2010}, who probe separations of $0.1 - 8\si{''}$, determined binary fractions of B and Be stars at $29\pm8$ and $30\pm8\%$. This finding could imply that the difference in the two binary fractions disappears towards larger separations.
            
            Exploiting the added information provided by Gaia, we performed a basic search for spatially resolved B and Be binaries by cross-matching the sample depicted in Fig. \ref{fig:results_summary}, and looking for sources within a radius of 10$\si{''}$. 
            Any `duplicate' sources that were retrieved were then deemed to be probable companions to the primary if they satisfied:

            \begin{equation} \label{eq:wide_bin}
                \lvert\omega_{primary} - \omega_{duplicate}\rvert \leq \sigma_{primary, \omega} + \sigma_{duplicate, \omega}
            \end{equation}
            
            where $\omega$ is the parallax of either star, and $\sigma_{\omega}$ is the uncertainty corresponding to either parallax.
            
            Based on this basic test, $6\pm1\%$ of B and $10\pm3\%$ of Be V-type stars are found to have potential resolved companions within a distance of $10\si{''}$. These values remained consistent even when randomly sampling or when generating semi-random samples of B stars intended to match the spectral type, parallax or apparent magnitude distributions of the Be stars. A summary of these results can also be found in Figure \ref{fig:results_summary}. In the case of each of the distribution matching samples, a Kolmogorov-Smirnov test was performed in order to verify that the distribution of each given parameter in the B star sample is likely being drawn from the same parent distribution as the Be star sample.
            
            Presuming that each of these additional sources is indeed a binary companion, poses an interesting question, that being why do Be stars have binary companions at a lower rate than their B type equivalents at close (i.e. $<0.7\si{''}$ separations, but at a similar rates at greater separations?

        \subsection {Further Constraints on Binary Separation} \label{sec:constraints}

            In this study we have determined, through the use of various astrometric measures, between separations of $0.02\,-\,0.7\si{''}$, the binary fraction of Be stars is significantly lower than that of their B star equivalents, while they are similar at larger separations. Let us continue by further identifying at which separation the B binary stars become more prevalent.

            \begin{table*}
                \centering
                \caption{Percentage of B and Be stars detected as binary at varying ranges of separations. By utilising combinations of the previously explored astrometric detection methods we are able to constrain the greater than $3\sigma$ death of Be binaries to between $0.02$ and $0.2\si{''}$.}
                \begin{tabular}{c|c|c|l}
                    \textbf{Separation Range / $\si{''}$} & \textbf{B} & \textbf{Be} & \textbf{Detection Method}          \\  \hline
                    $0.02 \leq x \leq 0.2$                & $28\pm1\%$ & $17\pm3\%$  & Detected by the Hipparcos-DR3      \\
                                                          &            &             & PMa but not the DR2-DR3 PMa.       \\
                    $0.02 \leq x \leq 0.7$                & $42\pm2\%$ & $28\pm4\%$  & Detected by the Hipparcos-DR3      \\
                                                          &            &             & PMa.                               \\
                    $0.2 \leq x \leq 1.1$                 & $27\pm1\%$ & $29\pm4\%$  & Detected by the Gaia DR2-DR3       \\
                                                          &            &             & PMa.                               \\
                    $0.04 \leq x \leq 0.7$                & $27\pm1\%$ & $20\pm4\%$  & Detected by the RUWE.              \\
                                                          &            &             &                                    \\
                    $0.1 \leq x \leq 8$                   & $29\pm8\%$ & $30\pm8\%$  & Detected by \cite{R_Oudmaijer_2010}      \\ 
                                                          &            &             & \\
                    $0.7 \leq x \leq 10$                  & $6\pm1\%$  & $10\pm3\%$  & Resolved as binary by Gaia.
                \end{tabular}
                \label{tab:separation_binary_frac}
            \end{table*}

            Table \ref{tab:separation_binary_frac} reveals that by comparing the two PMa values and looking at those systems detected by the Hipparcos-Gaia DR3 PMa but not by the Gaia DR2-DR3 PMa, we can constrain the lack of Be binaries to between $0.02$ and $0.2\si{''}$. We also note that at separations greater than $0.2\si{''}$, B and Be stars have similar fractions of their populations in binary systems - in agreement with the previous most recent comparative study of these stars \citep{R_Oudmaijer_2010} \footnote{Although we do not wish to make quantitative comparisons between values reported by our different detection methods, a larger binary fraction detected by the Hipparcos-Gaia DR3 PMa ($42\pm2\%$) than the Gaia DR2-DR3 PMa ($27\pm1\%$) would be expected in the case of a log-flat distribution of separations (as per \"{O}piks law; \citealt{E_Opik_1924}). This is because the separation range probed by the Hipparcos-Gaia DR3 PMa ($0.02$ to $0.7\si{''}$) is almost twice the width in log-space as that of the Gaia DR2-DR3 PMa ($0.2$ to $1.1\si{''}$). Conversely, by themselves the values reported for the Be stars may indicate a distribution decreasing towards to closer separations that we are sensitive to.}. Additionally, we note that the RUWE detects statistically similar rates of B and Be stars as being in binary systems. Thus, by virtue of the RUWE's sensitivity stretching between $0.04$ and $0.7\si{''}$, we are able to confine the dearth of Be binaries even further - restricting them to between $0.02$ and $0.04\si{''}$.

    \section{Discussion}\label{sec:discussion}

        In previous sections we have established, for the first time, ranges of angular separation and magnitude difference within which the PMa and RUWE are sensitive to binarity.
        
        We have then used these measures to reveal a greater than $3\sigma$ dearth of Be stars in binary systems at separations unresolved by Gaia, compared to B stars, we then go to further constrain this absence of Be binaries to separations between $0.02$ and $0.04\si{''}$.

        These results pose a conundrum, as one might naively assume that this implies that the Be phenomenon more favourably occurs via single star pathways, rather than the currently favoured binary pathways. Therefore, in order to continue to pursue the binary pathway for Be formation, we must consider: what sort of companion will remain undetected by these methods, whilst also remaining consistent with such models of Be formation?

        In this section we will posit some explanations for this difference in binary fraction, alongside the implication for models regarding the formation of the Be phenomenon.

        \subsection {Stripped Companions}

            One such solution invokes the presence of a companion whose envelope has been stripped by the putative Be star. Accretion from such a companion would provide a source for the initial spin-up mechanism that eventually results in the formation of the Be stars' characteristic disc \citep{K_ElBadry_2022, C_E_Jones_2022}, while also allowing the companion to become low-mass, dim or close enough to the host star \citep{A_J_Frost_2022} to be unlikely to be detected via the PMa. 
            
            Such a model has been suggested to be the route through which $20 - 100\%$ of field Be stars form \citep{K_ElBadry_2021b}, and a significant number of these stripped companions to Be stars have been reported in the literature (\citealt{K_ElBadry_2022, J_Bodensteiner_2020b, A_J_Frost_2022}; see Table \ref{tab:stripped_binaries}). Additionally, no close main sequence companions to Be stars have been confirmed \citep{J_Bodensteiner_2020}, which is consistent with what would be expected in the case of a stripped companion origin for Be stars (although it should be noted that this study only considers massive Be stars - i.e. spectral types earlier than B1.5).  As such, they may be responsible for hydrogen-poor supernovae and neutron star mergers resulting in gravitational wave sources.
        
            The periods proposed to be required in order for stripping to begin by \cite{Z_Han_2002} are of order $0.1 - 10$ days for Common Envelope Ejection and $400 - 1500$ days for Roche Lobe Overflow.  Using the average distance to the Be stars in our sample ($\sim 260\si{pc}$), and assuming a circular, Keplerian orbit, these stars would require a binary companion at a separation of $0.02\si{''}$ (approximately $5\,\si{au}$) or closer in order for stripping to initiate. While, the process of mass transfer would then act to widen the orbit of this companion, the resultant stripped star would be low-mass and faint. We would thus expect that the majority of stripped companions, if they indeed exist, would be missed by both our PMa values and the RUWE as they will too close to the host Be star, or too low mass and dim to be detected.
            
            In support of this, PMa and RUWE values were obtained for $16$ Be stars confirmed via X-ray observations to host a stripped companion (all of those from \citealt{Y_Naze_2022}). In order for the above suggestion to have any validity, these stars, that are known to host stripped companions, should not be detected as binaries using the measures detailed within this paper. A summary of these objects and their Gaia data can be found in Table \ref{tab:stripped_binaries}.

            \begin{table}
                \centering
                \caption{Summary of Gaia DR3 - Hipparcos PMa and RUWE values for Be stars quoted in \protect\cite{Y_Naze_2022} as being known to host stripped companions. PMa and RUWE values marked with an asterisk (*) constitute binary detections.}
                \begin{tabular}{c|l|c|c}
                     \textbf{Name}  & \textbf{Spectral Type} & \textbf{PMa} & \textbf{RUWE} \\ \hline
                     $\varphi$ Per  & B1.5Ve                 & $0.23$       & $1.804^*$ \\
                     HD 29441       & B2.5Vne                & $1.17$       & $1.125$ \\
                     HD 41335       & B1.5IV-Ve              & $8.13^*$     & $1.515^*$ \\
                     HD 43544       & B3Ve                   & $0.66$       & $1.047$ \\
                     HD 51354       & B3Ve                   & $0.79$       & $1.055$ \\
                     HD 58978       & B0.5IVe                & $2.52$       & $0.941$ \\
                     HD 60855       & B2Ve                   & $1.52$       & $1.097$ \\
                     HD 113120      & B2IVne                 & $6.93^*$     & $2.385^*$ \\
                     $\kappa^1$ Aps & B2Vnpe                 & $1.48$       & $1.060$ \\  
                     HD 152478      & B3Vnpe                 & $1.82$       & $0.990$ \\
                     HD 157832      & B1.5Ve                 & $1.60$       & $0.735$ \\
                     28 Cyg         & B3IVe                  & $1.55$       & $1.113$ \\
                     HD 194335      & B2IIIe                 & $3.81^*$     & $1.041$ \\
                     59 Cyg         & B1Ve                   & $6.38^*$     & $2.724^*$ \\
                     60 Cyg         & B1Ve                   & $6.31^*$     & $0.142$ \\
                     8 Lac A        & B1IVe                  & $0.87$       & $1.480^*$ \\
                \end{tabular} 
                \label{tab:stripped_binaries}
            \end{table}

            Seven of these 16 systems are detected by either method; 5 by the Hipparcos - Gaia DR3 PMa, and 5 by the RUWE, with 3 objects detected by both. However, 5 of these 7 detected systems (the exceptions being HD 194335 and HD 200310) have known additional companions \citep{WDS} at separations within the range the PMa's and RUWE are sensitive to. In addition, these companions are at separations significantly larger than required for the periods reported by \cite{Y_Naze_2022}, and are therefore inconsistent with being the reported stripped companion. An undetected, third, companion has also been suggested in the case of HD 200310 to explain the eccentric orbit of the stripped companion \citep{R_Klement_2022}, making the system a potential triple. Thus, the true fraction of detected stripped companions would seem to be either $6$ or $12\%$ depending on how one wishes to treat the case of HD 200310. Therefore, we reach the conclusion that the stripped companions listed in \cite{Y_Naze_2022} are by and large not detected by the PMa or RUWE, whose detections turn out to trace a third body in these systems instead. Thus, the potential for the Be phenomenon being caused by the presence of stripped companions remains consistent with our results.

        \subsection {Triples}

            Intriguingly, the separation range of the ``missing'' Be binaries (0.02 - 0.04$\si{''}$; with 0.02$\si{''}$ corresponding to 5$\si{au}$) is typically too large for mass transfer to have occurred (typically less than $4\si{au}$ is required per the periods reported in \citealt{Z_Han_2002}).  The natural next question to ask is then, why would the Be stars lack binaries in this particular separation range, while the stripping of a close companion should occur at smaller separations?
            
            A clue can be found in the fact that the systems listed in Table \ref{tab:stripped_binaries} show a high incidence rate of higher order multiplicity, with $56\pm18\%$ of this sample of Be stars known to host a stripped companion having additional companions \citep{WDS} at separations larger than those reported in \cite{Y_Naze_2022}. Additionally this number is also somewhat elevated with the expected incidence rate of higher order multiplicity posited by \cite{M_Moe_2017} - being around $40\%$.
            
            The presence of such a third companion can neatly explain the absence of detected Be binaries at larger separations, while also allowing for any companion stars to become close enough for mass transfer to occur; it is well known that higher order multiplicity can result in the hardening of an inner binary. Indeed, a third body increases the occurrences of migration and eventual binary interactions significantly \citep{S_Toonen_2022, Kummer_F_2023}.

            Recent models \citep{H_P_Preece_2022} suggest that of the systems that eventually form stripped companions - half form via binary pathways and half form via triple pathways. This can take place by either mass transfer occurring within the inner binary - with the outer tertiary star potentially becoming unbound - or a merger of the inner binary and mass transfer occurring between the two remaining stars \citep{H_P_Preece_2022}.
            
            Our results therefore imply that close binary interactions are responsible for the formation of Be stars, which constitute $20\%$ of the B star population, with a migratory effect causing the lack of Be binaries in the $0.02$ to $0.04\si{''}$. Moreover, we suggest that triplicity must play a vital role in triggering this migration, and thus in the formation of Be stars as a whole.

        \subsection {Potential Limitations of the Methods}
            
            We will now address the various potential limitations of the methods as described in Sections \ref{sec:methods} and \ref{sec:results}.

            \subsubsection {Sample Selection}

                First, we note that the Be phenomenon is known to be transient, with the disc known to disappear and reappear on short timescales, which can lead to the potential misclassification of Be stars as B stars \cite{T_Rivinius_2013}. Thus, contamination of the B star sample with misclassified Be stars could effect the previously detected binary fractions for the B star sample. However, as the Be sample should have low contamination (as observation of the emission line features associated with the Be phenomenon is required for classification as a Be star, even if these features later disappear), the detected Be binary fractions should be representative of the Be population as a whole - and thus representative of any misclassified Be stars. Therefore, should there be a significant amount of Be contaminants in the B star sample, then they will act to bring the true B binary fraction down - thus the true B binary fraction would differ more significantly from the Be fraction.

                We also note that, as this sample is selected from the BSC \citep{BSC_book, BSC, BSC_Supplement}, these stars are necessarily bright. Whilst Gaia astrometric fits become systematic dominated at $G < 13$ \citep{L_Lindegren_2020}, the resultant increase in uncertainties associated with this effect ensures that the PMa is still a viable metric - as a signal-to-noise ratio is used to quantify binary detection. Additionally, in the calculation of the new Gaia DR2 - DR3 PMa's corrections to the respective proper motions of bright stars as proscribed by \cite{L_Lindegren_2018b} and \cite{T_Cantat-Gaudin_2021} have been applied. Steps have also been taken to ensure that the two samples are subject to the same selection effects, and these are detailed in Section \ref{sec:biases}, thus any systematic effects should be consistent between the two samples. To check this, all stars brighter than $5\si{mag}$ were removed from the B and Be samples; which resulted in zero change to the reported dearth of Be binaries within the $0.02\,-\,0.2\si{''}$ range.

            \subsubsection {Variability}
            
                While photometric variability has not been quantitatively considered here, we anticipate that it is unlikely to effect the PMa unless the variability is occurring both asymmetrically and on a scale similar to the sensitive separation range, here we look at this in more detail.
                
                First, consider a single star exhibiting symmetrical variation; in this case the position of the photocentre does not deviate from the centre of mass - thus leading to no proper motion anomaly. When we then introduce an unresolved companion, the position of the photocentre therefore deviates from the position of the centre of mass, however the magnitude of the deviation varies with the variability cycle. Thus, while such variability shouldn't produce a false positive detection of binarity, it is possible for it to induce a false negative. The likelihood of this false negative thus depend on two factors: the magnitude of the variability, and the timescale on which it occurs (as observations taken at the same point in the variability cycle will obviously be unaffected, and variability with a period shorter than either individual observing period will be smeared, diminishing the effect on the PMa).
                
                In the case of asymmetric variability - i.e. that caused by phenomena such as mass ejection, or variations in surface brightness in stars such as red super-giants \citep{A_Chiavassa_2022}, or asymmetry within a disc \citep{A_Meilland_2007, Ph_Stee_2013} - deviation of the photocentre from the centre of mass will occur. However, such asymmetry would have to be on a similar scale to the PMa's sensitive range to in order for such deviation to be construed as a false positive detection of binarity.

                In the specific case of this study, we anticipate little effect of the photometric variability of Be stars \citep{J_Labadie-Bartz_2017, J_Labadie-Bartz_2022} on our PMa values. This is due to the observed periods of Be variability being significantly shorter than the observing periods of Gaia DR2, DR3 and Hipparcos ($668$, $1038$ and $1227$ days respectively), with the longest period variability found by \cite{J_Labadie-Bartz_2017} being approximately $200$ days. Similarly, variability caused by mass loss onto the Be disc is unlikely to significantly effect the PMa, as the discs of Be stars, with sizes of order milli-arcseconds, are very small when compared to the resolution achieved by Gaia DR3 (i.e. $0.7\si{''}$; \cite{H_Wheelwright_2012, A_Quirrenbach_1997}). Additionally, by similar logic, asymmetry of the discs themselves \citep{A_Meilland_2007, Ph_Stee_2013} is also expected to have little effect on the PMa measurements of these stars. Likewise, the stars are necessariliy smaller than their disc, they are expected to be too small to have asymmetric surface brightness that would effect the PMa.

    \section {Conclusions}

        In this paper we have performed the largest scale comparative study of B and Be-star binarity to date, utilising both the Hipparcos - Gaia DR3 PMa values found in \cite{P_Kervella_2021} and new Gaia DR2 - DR3 PMa values calculated for the first time here, alongside the Gaia provided RUWE. 
        
        We evaluate the limits in which these methods are sensitive to binarity.  Particularly we determine for the first time the range of angular separations they are capable of reliably detecting, through the use of WDS \citep{WDS} data. We thus find that the Hipparcos - Gaia DR3 and Gaia DR2 - DR3 PMa's are sensitive to binaries with separations in the $0.02\,-\,0.7\si{''}$ and $0.2\,-\,1.1\si{''}$ ranges respectively. Likewise the RUWE is found to be sensitive to separations between $0.04$ and $0.7\si{''}$.

        Through the use of these astrometric methods in combination with one another, we report a greater than $3\sigma$ dearth of Be binaries compared to B stars at separations between $0.02$ and $0.2\si{''}$, with the fractions of B and Be stars found to be in binary systems at these separations being $28\pm1$ and $17\pm3\%$ respectively.

        We posit that this apparent lack of Be binaries provides evidence for the presence of stripped companions to the Be stars, with the mass-transfer process by which they are created having been previously suggested as an origin for the Be phenomenon. Such stars would go undetected by the measures used within this paper, as they are too close, low-mass and dim to induce a significant enough deviation of the photocentre from the centre of mass. We also find that Be stars previously reported to have a stripped companion have additional companions (i.e. triples and higher order multiples) at a rate somewhat elevated from the expected rate for B stars as a whole. Therefore, we suggest that migration of binary companions to Be stars to separations at which mass-transfer can occur, triggered by a third companion, must play an important role in the formation of Be stars.
        
    \section * {Acknowledgements}
        
        The STARRY project has received funding from the European Union’s Horizon 2020 research and innovation programme under MSCA ITN-EID grant agreement No 676036. This project has also received funding from the European Union's Framework Programme for Research and Innovation Horizon 2020 (2014-2020) under the Marie Sk\l{}odowska-Curie Grant Agreement No. 823734. This work has made use of data from the European Space Agency (ESA) mission {\it Gaia }(\url{https://www.cosmos.esa.int/gaia}), processed by the {\it Gaia} Data Processing and Analysis Consortium (DPAC, \url{https://www.cosmos.esa.int/web/gaia/dpac/consortium}). Funding for the DPAC has been provided by national institutions, in particular the institutions participating in the {\it Gaia} Multilateral Agreement. 

    \section * {Data Availability}

        The catalogues of the proper motion anomalies and RUWE values of both the sample of B and Be stars, as well as the WDS binaries are available form the corresponding author on reasonable request, and will also be made available via VizieR.

    \bibliographystyle{mnras}
    \bibliography{final} 

\newcommand{\noop}[1]{}
\begin{thebibliography}{}
\makeatletter
\relax
\def\mn@urlcharsother{\let\do\@makeother \do\$\do\&\do\#\do\^\do\_\do\%\do\~}
\def\mn@doi{\begingroup\mn@urlcharsother \@ifnextchar [ {\mn@doi@}
  {\mn@doi@[]}}
\def\mn@doi@[#1]#2{\def\@tempa{#1}\ifx\@tempa\@empty \href
  {http://dx.doi.org/#2} {doi:#2}\else \href {http://dx.doi.org/#2} {#1}\fi
  \endgroup}
\def\mn@eprint#1#2{\mn@eprint@#1:#2::\@nil}
\def\mn@eprint@arXiv#1{\href {http://arxiv.org/abs/#1} {{\tt arXiv:#1}}}
\def\mn@eprint@dblp#1{\href {http://dblp.uni-trier.de/rec/bibtex/#1.xml}
  {dblp:#1}}
\def\mn@eprint@#1:#2:#3:#4\@nil{\def\@tempa {#1}\def\@tempb {#2}\def\@tempc
  {#3}\ifx \@tempc \@empty \let \@tempc \@tempb \let \@tempb \@tempa \fi \ifx
  \@tempb \@empty \def\@tempb {arXiv}\fi \@ifundefined
  {mn@eprint@\@tempb}{\@tempb:\@tempc}{\expandafter \expandafter \csname
  mn@eprint@\@tempb\endcsname \expandafter{\@tempc}}}

\bibitem[\protect\citeauthoryear{{Abt} \& {Cardona}}{{Abt} \&
  {Cardona}}{1984}]{H_A_Abt_1984}
{Abt} H.~A.,  {Cardona} O.,  1984, \mn@doi [The Astrophysical Journal]
  {10.1086/162490}, \href
  {https://ui.adsabs.harvard.edu/abs/1984ApJ...285..190A} {285, 190}

\bibitem[\protect\citeauthoryear{Abt \& Levy}{Abt \& Levy}{1978}]{H_A_Abt_1978}
Abt H.~A.,  Levy S.~G.,  1978, The Astrophysical Journal Supplement Series, 36,
  241

\bibitem[\protect\citeauthoryear{{Belokurov} et~al.,}{{Belokurov}
  et~al.}{2020}]{V_Belokurov_2020}
{Belokurov} V.,  et~al., 2020, \mn@doi [Monthly Notices of the Royal
  Astronomical Society] {10.1093/mnras/staa1522}, \href
  {https://ui.adsabs.harvard.edu/abs/2020MNRAS.496.1922B} {496, 1922}

\bibitem[\protect\citeauthoryear{{Bodenheimer}}{{Bodenheimer}}{1995}]{P_Bodenheimer_1995}
{Bodenheimer} P.,  1995, \mn@doi [Annual Review of Astronomy and Astrophysics]
  {10.1146/annurev.aa.33.090195.001215}, \href
  {https://ui.adsabs.harvard.edu/abs/1995ARA&A..33..199B} {33, 199}

\bibitem[\protect\citeauthoryear{{Bodensteiner}, {Shenar}  \&
  {Sana}}{{Bodensteiner} et~al.}{2020a}]{J_Bodensteiner_2020}
{Bodensteiner} J.,  {Shenar} T.,   {Sana} H.,  2020a, \mn@doi [Astronomy \&
  Astrophysics] {10.1051/0004-6361/202037640}, \href
  {https://ui.adsabs.harvard.edu/abs/2020A&A...641A..42B} {641, A42}

\bibitem[\protect\citeauthoryear{{Bodensteiner} et~al.,}{{Bodensteiner}
  et~al.}{2020b}]{J_Bodensteiner_2020b}
{Bodensteiner} J.,  et~al., 2020b, \mn@doi [Astronomy \& Astrophysics]
  {10.1051/0004-6361/202038682}, \href
  {https://ui.adsabs.harvard.edu/abs/2020A&A...641A..43B} {641, A43}

\bibitem[\protect\citeauthoryear{{Boubert} \& {Evans}}{{Boubert} \&
  {Evans}}{2018}]{D_Boubert_2018}
{Boubert} D.,  {Evans} N.~W.,  2018, \mn@doi [Monthly Notices of the Royal
  Astronomical Society] {10.1093/mnras/sty980}, \href
  {https://ui.adsabs.harvard.edu/abs/2018MNRAS.477.5261B} {477, 5261}

\bibitem[\protect\citeauthoryear{{Brandt}}{{Brandt}}{2018}]{T_Brandt_2018}
{Brandt} T.~D.,  2018, \mn@doi [The Astrophysical Journal Supplement Series]
  {10.3847/1538-4365/aaec06}, \href
  {https://ui.adsabs.harvard.edu/abs/2018ApJS..239...31B} {239, 31}

\bibitem[\protect\citeauthoryear{{Brandt}}{{Brandt}}{2021}]{T_Brandt_2021}
{Brandt} T.~D.,  2021, \mn@doi [\apjs] {10.3847/1538-4365/abf93c}, \href
  {https://ui.adsabs.harvard.edu/abs/2021ApJS..254...42B} {254, 42}

\bibitem[\protect\citeauthoryear{{Cantat-Gaudin} \& {Brandt}}{{Cantat-Gaudin}
  \& {Brandt}}{2021}]{T_Cantat-Gaudin_2021}
{Cantat-Gaudin} T.,  {Brandt} T.~D.,  2021, \mn@doi [Astronomy \& Astrophysics]
  {10.1051/0004-6361/202140807}, \href
  {https://ui.adsabs.harvard.edu/abs/2021A&A...649A.124C} {649, A124}

\bibitem[\protect\citeauthoryear{{Chiavassa}, {Kudritzki}, {Davies}, {Freytag}
  \& {de Mink}}{{Chiavassa} et~al.}{2022}]{A_Chiavassa_2022}
{Chiavassa} A.,  {Kudritzki} R.,  {Davies} B.,  {Freytag} B.,   {de Mink}
  S.~E.,  2022, \mn@doi [Astronomy \& Astrophysics]
  {10.1051/0004-6361/202243568}, \href
  {https://ui.adsabs.harvard.edu/abs/2022A&A...661L...1C} {661, L1}

\bibitem[\protect\citeauthoryear{{Draper}, {Wisniewski}, {Bjorkman}, {Meade},
  {Haubois}, {Mota}, {Carciofi}  \& {Bjorkman}}{{Draper}
  et~al.}{2014}]{Z_H_Draper_2014}
{Draper} Z.~H.,  {Wisniewski} J.~P.,  {Bjorkman} K.~S.,  {Meade} M.~R.,
  {Haubois} X.,  {Mota} B.~C.,  {Carciofi} A.~C.,   {Bjorkman} J.~E.,  2014,
  \mn@doi [The Astrophysical Journal] {10.1088/0004-637X/786/2/120}, \href
  {https://ui.adsabs.harvard.edu/abs/2014ApJ...786..120D} {786, 120}

\bibitem[\protect\citeauthoryear{{El-Badry} \& {Quataert}}{{El-Badry} \&
  {Quataert}}{2021}]{K_ElBadry_2021b}
{El-Badry} K.,  {Quataert} E.,  2021, \mn@doi [Monthly Notices of the Royal
  Astronomical Society] {10.1093/mnras/stab285}, \href
  {https://ui.adsabs.harvard.edu/abs/2021MNRAS.502.3436E} {502, 3436}

\bibitem[\protect\citeauthoryear{{El-Badry}, {Rix}  \& {Heintz}}{{El-Badry}
  et~al.}{2021}]{K_El_Badry_2021}
{El-Badry} K.,  {Rix} H.-W.,   {Heintz} T.~M.,  2021, \mn@doi [Monthly Notices
  of the Royal Astronomical Society] {10.1093/mnras/stab323}, \href
  {https://ui.adsabs.harvard.edu/abs/2021MNRAS.506.2269E} {506, 2269}

\bibitem[\protect\citeauthoryear{{El-Badry} et~al.,}{{El-Badry}
  et~al.}{2022}]{K_ElBadry_2022}
{El-Badry} K.,  et~al., 2022, \mn@doi [Monthly Notices of the Royal
  Astronomical Society] {10.1093/mnras/stac2422}, \href
  {https://ui.adsabs.harvard.edu/abs/2022MNRAS.516.3602E} {516, 3602}

\bibitem[\protect\citeauthoryear{{Frost} et~al.,}{{Frost}
  et~al.}{2022}]{A_J_Frost_2022}
{Frost} A.~J.,  et~al., 2022, \mn@doi [Astronomy \& Astrophysics]
  {10.1051/0004-6361/202143004}, \href
  {https://ui.adsabs.harvard.edu/abs/2022A&A...659L...3F} {659, L3}

\bibitem[\protect\citeauthoryear{{Gaia Collaboration} et~al.,}{{Gaia
  Collaboration} et~al.}{2018}]{Gaia_Collaboration_2018}
{Gaia Collaboration} et~al., 2018, \mn@doi [Astronomy \& Astrophysics]
  {10.1051/0004-6361/201833051}, \href
  {https://ui.adsabs.harvard.edu/abs/2018A&A...616A...1G} {616, A1}

\bibitem[\protect\citeauthoryear{{Gaia Collaboration} et~al.,}{{Gaia
  Collaboration} et~al.}{2021}]{Gaia_Collaboration_2021}
{Gaia Collaboration} et~al., 2021, \mn@doi [Astronomy and Astrophysics]
  {10.1051/0004-6361/202039657}, \href
  {https://ui.adsabs.harvard.edu/abs/2021A&A...649A...1G} {649, A1}

\bibitem[\protect\citeauthoryear{{Gaia Collaboration} et~al.,}{{Gaia
  Collaboration} et~al.}{2022}]{Gaia_Collaboration_2022}
{Gaia Collaboration} et~al., 2022, arXiv e-prints, \href
  {https://ui.adsabs.harvard.edu/abs/2022arXiv220800211G} {p. arXiv:2208.00211}

\bibitem[\protect\citeauthoryear{{Granada}, {Ekstr{\"o}m}, {Georgy},
  {Krti{\v{c}}ka}, {Owocki}, {Meynet}  \& {Maeder}}{{Granada}
  et~al.}{2013}]{A_Granada_2013}
{Granada} A.,  {Ekstr{\"o}m} S.,  {Georgy} C.,  {Krti{\v{c}}ka} J.,  {Owocki}
  S.,  {Meynet} G.,   {Maeder} A.,  2013, \mn@doi [Astronomy \& Astrophysics]
  {10.1051/0004-6361/201220559}, \href
  {https://ui.adsabs.harvard.edu/abs/2013A&A...553A..25G} {553, A25}

\bibitem[\protect\citeauthoryear{{Griffiths}, {Hicks}  \& {Milone}}{{Griffiths}
  et~al.}{1988}]{S_C_Griffiths_1988}
{Griffiths} S.~C.,  {Hicks} R.~B.,   {Milone} E.~F.,  1988, Journal of the
  Royal Astronomical Society of Canada, \href
  {https://ui.adsabs.harvard.edu/abs/1988JRASC..82....1G} {82, 1}

\bibitem[\protect\citeauthoryear{{Han}, {Podsiadlowski}, {Maxted}, {Marsh}  \&
  {Ivanova}}{{Han} et~al.}{2002}]{Z_Han_2002}
{Han} Z.,  {Podsiadlowski} P.,  {Maxted} P.~F.~L.,  {Marsh} T.~R.,   {Ivanova}
  N.,  2002, \mn@doi [Monthly Notices of the Royal Astronomical Society]
  {10.1046/j.1365-8711.2002.05752.x}, \href
  {https://ui.adsabs.harvard.edu/abs/2002MNRAS.336..449H} {336, 449}

\bibitem[\protect\citeauthoryear{{Hastings}, {Wang}  \& {Langer}}{{Hastings}
  et~al.}{2020}]{B_Hastings_2020}
{Hastings} B.,  {Wang} C.,   {Langer} N.,  2020, \mn@doi [Astronomy \&
  Astrophysics] {10.1051/0004-6361/201937018}, \href
  {https://ui.adsabs.harvard.edu/abs/2020A&A...633A.165H} {633, A165}

\bibitem[\protect\citeauthoryear{{Hastings}, {Langer}, {Wang}, {Schootemeijer}
  \& {Milone}}{{Hastings} et~al.}{2021}]{B_Hastings_2021}
{Hastings} B.,  {Langer} N.,  {Wang} C.,  {Schootemeijer} A.,   {Milone} A.~P.,
   2021, \mn@doi [Astronomy \& Astrophysics] {10.1051/0004-6361/202141269},
  \href {https://ui.adsabs.harvard.edu/abs/2021A&A...653A.144H} {653, A144}

\bibitem[\protect\citeauthoryear{{Hoffleit} \& {Jaschek}}{{Hoffleit} \&
  {Jaschek}}{1982}]{BSC_book}
{Hoffleit} D.,  {Jaschek} C.,  1982, {The Bright Star Catalogue. Fourth revised
  edition. (Containing data compiled through 1979).}.
New Haven: Yale University Observatory

\bibitem[\protect\citeauthoryear{{Hoffleit} \& {Saladyga}}{{Hoffleit} \&
  {Saladyga}}{1997}]{BSC_Supplement}
{Hoffleit} D.,  {Saladyga} M.,  1997, VizieR Online Data Catalog, \href
  {https://ui.adsabs.harvard.edu/abs/1997yCat.5036....0H} {p. V/36B}

\bibitem[\protect\citeauthoryear{{Hoffleit} \& {Warren}}{{Hoffleit} \&
  {Warren}}{1995}]{BSC}
{Hoffleit} D.,  {Warren} W.~H. J.,  1995, VizieR Online Data Catalog, \href
  {https://ui.adsabs.harvard.edu/abs/1995yCat.5050....0H} {p.~V/50}

\bibitem[\protect\citeauthoryear{{Jones} et~al.,}{{Jones}
  et~al.}{2022}]{C_E_Jones_2022}
{Jones} C.~E.,  et~al., 2022, \mn@doi [Astrophysics and Space Science]
  {10.1007/s10509-022-04127-5}, \href
  {https://ui.adsabs.harvard.edu/abs/2022Ap&SS.367..124J} {367, 124}

\bibitem[\protect\citeauthoryear{{Kervella}, {Arenou}, {Mignard}  \&
  {Th{\'e}venin}}{{Kervella} et~al.}{2019}]{P_Kervella_2019a}
{Kervella} P.,  {Arenou} F.,  {Mignard} F.,   {Th{\'e}venin} F.,  2019, \mn@doi
  [Astronomy \& Astrophysics] {10.1051/0004-6361/201834371}, \href
  {https://ui.adsabs.harvard.edu/abs/2019A&A...623A..72K} {623, A72}

\bibitem[\protect\citeauthoryear{{Kervella}, {Arenou}  \&
  {Th{\'e}venin}}{{Kervella} et~al.}{2022}]{P_Kervella_2021}
{Kervella} P.,  {Arenou} F.,   {Th{\'e}venin} F.,  2022, \mn@doi [Astronomy \&
  Astrophysics] {10.1051/0004-6361/202142146}, \href
  {https://ui.adsabs.harvard.edu/abs/2022A&A...657A...7K} {657, A7}

\bibitem[\protect\citeauthoryear{{Klement} et~al.,}{{Klement}
  et~al.}{2019}]{R_Klement_2019}
{Klement} R.,  et~al., 2019, \mn@doi [The Astrophysical Journal]
  {10.3847/1538-4357/ab48e7}, \href
  {https://ui.adsabs.harvard.edu/abs/2019ApJ...885..147K} {885, 147}

\bibitem[\protect\citeauthoryear{{Klement} et~al.,}{{Klement}
  et~al.}{2022}]{R_Klement_2022}
{Klement} R.,  et~al., 2022, \mn@doi [The Astrophysical Journal]
  {10.3847/1538-4357/ac4266}, \href
  {https://ui.adsabs.harvard.edu/abs/2022ApJ...926..213K} {926, 213}

\bibitem[\protect\citeauthoryear{{Kummer}, {Toonen}  \& {de Koter}}{{Kummer}
  et~al.}{2023}]{Kummer_F_2023}
{Kummer} F.,  {Toonen} S.,   {de Koter} A.,  2023, \mn@doi [arXiv e-prints]
  {10.48550/arXiv.2306.09400}, \href
  {https://ui.adsabs.harvard.edu/abs/2023arXiv230609400K} {p. arXiv:2306.09400}

\bibitem[\protect\citeauthoryear{{Labadie-Bartz} et~al.,}{{Labadie-Bartz}
  et~al.}{2017}]{J_Labadie-Bartz_2017}
{Labadie-Bartz} J.,  et~al., 2017, \mn@doi [The Astronomical Journal]
  {10.3847/1538-3881/aa6396}, \href
  {https://ui.adsabs.harvard.edu/abs/2017AJ....153..252L} {153, 252}

\bibitem[\protect\citeauthoryear{{Labadie-Bartz}, {Carciofi}, {Henrique de
  Amorim}, {Rubio}, {Luiz Figueiredo}, {Ticiani dos Santos}  \&
  {Thomson-Paressant}}{{Labadie-Bartz} et~al.}{2022}]{J_Labadie-Bartz_2022}
{Labadie-Bartz} J.,  {Carciofi} A.~C.,  {Henrique de Amorim} T.,  {Rubio} A.,
  {Luiz Figueiredo} A.,  {Ticiani dos Santos} P.,   {Thomson-Paressant} K.,
  2022, \mn@doi [The Astronomical Journal] {10.3847/1538-3881/ac5abd}, \href
  {https://ui.adsabs.harvard.edu/abs/2022AJ....163..226L} {163, 226}

\bibitem[\protect\citeauthoryear{{Lindegren}}{{Lindegren}}{2020}]{L_Lindegren_2020}
{Lindegren} L.,  2020, \mn@doi [Astronomy \& Astrophysics]
  {10.1051/0004-6361/201936161}, \href
  {https://ui.adsabs.harvard.edu/abs/2020A&A...633A...1L} {633, A1}

\bibitem[\protect\citeauthoryear{{Lindegren} et~al.,}{{Lindegren}
  et~al.}{2018}]{L_Lindegren_2018b}
{Lindegren} L.,  et~al., 2018, \mn@doi [Astronomy \& Astrophysics]
  {10.1051/0004-6361/201832727}, \href
  {https://ui.adsabs.harvard.edu/abs/2018A&A...616A...2L} {616, A2}

\bibitem[\protect\citeauthoryear{{Lindegren} et~al.,}{{Lindegren}
  et~al.}{2021}]{L_Lindegren_2021}
{Lindegren} L.,  et~al., 2021, \mn@doi [Astronomy and Astrophysics]
  {10.1051/0004-6361/202039709}, \href
  {https://ui.adsabs.harvard.edu/abs/2021A&A...649A...2L} {649, A2}

\bibitem[\protect\citeauthoryear{{Mason}, {Wycoff}, {Hartkopf}, {Douglass}  \&
  {Worley}}{{Mason} et~al.}{2022}]{WDS}
{Mason} B.~D.,  {Wycoff} G.~L.,  {Hartkopf} W.~I.,  {Douglass} G.~G.,
  {Worley} C.~E.,  2022, VizieR Online Data Catalog, \href
  {https://ui.adsabs.harvard.edu/abs/2022yCat....102026M} {p. B/wds}

\bibitem[\protect\citeauthoryear{{Meilland} et~al.,}{{Meilland}
  et~al.}{2007}]{A_Meilland_2007}
{Meilland} A.,  et~al., 2007, \mn@doi [Astronomy \& Astrophysics]
  {10.1051/0004-6361:20065410}, \href
  {https://ui.adsabs.harvard.edu/abs/2007A&A...464...73M} {464, 73}

\bibitem[\protect\citeauthoryear{{Moe} \& {Di Stefano}}{{Moe} \& {Di
  Stefano}}{2017}]{M_Moe_2017}
{Moe} M.,  {Di Stefano} R.,  2017, \mn@doi [The Astrophysical Journal
  Supplement] {10.3847/1538-4365/aa6fb6}, \href
  {https://ui.adsabs.harvard.edu/abs/2017ApJS..230...15M} {230, 15}

\bibitem[\protect\citeauthoryear{{Naz{\'e}}, {Rauw}, {Smith}  \&
  {Motch}}{{Naz{\'e}} et~al.}{2022}]{Y_Naze_2022}
{Naz{\'e}} Y.,  {Rauw} G.,  {Smith} M.~A.,   {Motch} C.,  2022, \mn@doi
  [Monthly Noticies of the Royal Astronomical Society]
  {10.1093/mnras/stac2245}, \href
  {https://ui.adsabs.harvard.edu/abs/2022MNRAS.516.3366N} {516, 3366}

\bibitem[\protect\citeauthoryear{{{\"O}pik}}{{{\"O}pik}}{1924}]{E_Opik_1924}
{{\"O}pik} E.,  1924, Publications of the Tartu Astrofizica Observatory, \href
  {https://ui.adsabs.harvard.edu/abs/1924PTarO..25f...1O} {25, 1}

\bibitem[\protect\citeauthoryear{{Oudmaijer} \& {Parr}}{{Oudmaijer} \&
  {Parr}}{2010}]{R_Oudmaijer_2010}
{Oudmaijer} R.~D.,  {Parr} A.~M.,  2010, \mn@doi [Monthly Notices of the Royal
  Astronomical Society] {10.1111/j.1365-2966.2010.16609.x}, \href
  {https://ui.adsabs.harvard.edu/abs/2010MNRAS.405.2439O} {405, 2439}

\bibitem[\protect\citeauthoryear{{Oudmaijer}, {Jones}  \& {Vioque}}{{Oudmaijer}
  et~al.}{2022}]{R_Oudmaijer_2022}
{Oudmaijer} R.~D.,  {Jones} E. R.~M.,   {Vioque} M.,  2022, \mn@doi [Monthly
  Notices of the Royal Astronomical Society] {10.1093/mnrasl/slac088}, \href
  {https://ui.adsabs.harvard.edu/abs/2022MNRAS.516L..61O} {516, L61}

\bibitem[\protect\citeauthoryear{Owocki \& Cranmer}{Owocki \&
  Cranmer}{2002}]{S_P_Owocki_2002}
Owocki S.,  Cranmer S.,  2002, \mn@doi [International Astronomical Union
  Colloquium] {10.1017/S0252921100016961}, 185, 512–519

\bibitem[\protect\citeauthoryear{{Penoyre}, {Belokurov}  \& {Evans}}{{Penoyre}
  et~al.}{2022}]{Z_Penoyre_2021}
{Penoyre} Z.,  {Belokurov} V.,   {Evans} N.~W.,  2022, \mn@doi [Monthly
  Noticies of the Royal Astronomical Society] {10.1093/mnras/stac959}, \href
  {https://ui.adsabs.harvard.edu/abs/2022MNRAS.513.2437P} {513, 2437}

\bibitem[\protect\citeauthoryear{{Pineau}, {Boch}, {Derri{\`e}re}  \&
  {Schaaff}}{{Pineau} et~al.}{2020}]{F_Pineau_2020}
{Pineau} F.-X.,  {Boch} T.,  {Derri{\`e}re} S.,   {Schaaff} A.,  2020, in
  {Ballester} P.,  {Ibsen} J.,  {Solar} M.,   {Shortridge} K.,  eds,
  Astronomical Society of the Pacific Conference Series Vol. 522, Astronomical
  Data Analysis Software and Systems XXVII. p.~125

\bibitem[\protect\citeauthoryear{{Porter} \& {Rivinius}}{{Porter} \&
  {Rivinius}}{2003}]{J_Porter_2003}
{Porter} J.~M.,  {Rivinius} T.,  2003, \mn@doi [Publications of the
  Astronomical Society of the Pacific] {10.1086/378307}, \href
  {https://ui.adsabs.harvard.edu/abs/2003PASP..115.1153P} {115, 1153}

\bibitem[\protect\citeauthoryear{{Preece}, {Hamers}, {Battich}  \&
  {Rajamuthukumar}}{{Preece} et~al.}{2022}]{H_P_Preece_2022}
{Preece} H.~P.,  {Hamers} A.~S.,  {Battich} T.,   {Rajamuthukumar} A.~S.,
  2022, \mn@doi [Monthly Notices of the Royal Astronomical Society]
  {10.1093/mnras/stac2798}, \href
  {https://ui.adsabs.harvard.edu/abs/2022MNRAS.517.2111P} {517, 2111}

\bibitem[\protect\citeauthoryear{{Quirrenbach} et~al.,}{{Quirrenbach}
  et~al.}{1997}]{A_Quirrenbach_1997}
{Quirrenbach} A.,  et~al., 1997, \mn@doi [The Astrophysical Journal]
  {10.1086/303854}, \href
  {https://ui.adsabs.harvard.edu/abs/1997ApJ...479..477Q} {479, 477}

\bibitem[\protect\citeauthoryear{{Rivinius}, {Carciofi}  \&
  {Martayan}}{{Rivinius} et~al.}{2013}]{T_Rivinius_2013}
{Rivinius} T.,  {Carciofi} A.~C.,   {Martayan} C.,  2013, \mn@doi [Astronomy \&
  Astrophysics Review] {10.1007/s00159-013-0069-0}, \href
  {https://ui.adsabs.harvard.edu/abs/2013A&ARv..21...69R} {21, 69}

\bibitem[\protect\citeauthoryear{{Semaan}, {Hubert}, {Zorec},
  {Guti{\'e}rrez-Soto}, {Fr{\'e}mat}, {Martayan}, {Fabregat}  \&
  {Eggenberger}}{{Semaan} et~al.}{2018}]{T_Semaan_2018}
{Semaan} T.,  {Hubert} A.~M.,  {Zorec} J.,  {Guti{\'e}rrez-Soto} J.,
  {Fr{\'e}mat} Y.,  {Martayan} C.,  {Fabregat} J.,   {Eggenberger} P.,  2018,
  \mn@doi [Astronomy \& Astrophysics] {10.1051/0004-6361/201629243}, \href
  {https://ui.adsabs.harvard.edu/abs/2018A&A...613A..70S} {613, A70}

\bibitem[\protect\citeauthoryear{{Shao} \& {Li}}{{Shao} \&
  {Li}}{2014}]{Y_Shao_2014}
{Shao} Y.,  {Li} X.-D.,  2014, \mn@doi [The Astrophysical Journal]
  {10.1088/0004-637X/796/1/37}, \href
  {https://ui.adsabs.harvard.edu/abs/2014ApJ...796...37S} {796, 37}

\bibitem[\protect\citeauthoryear{{Stassun} \& {Torres}}{{Stassun} \&
  {Torres}}{2021}]{K_Stassun_2021}
{Stassun} K.~G.,  {Torres} G.,  2021, \mn@doi [The Astrophysical Journal
  Letters] {10.3847/2041-8213/abdaad}, \href
  {https://ui.adsabs.harvard.edu/abs/2021ApJ...907L..33S} {907, L33}

\bibitem[\protect\citeauthoryear{{Stee} et~al.,}{{Stee}
  et~al.}{2013}]{Ph_Stee_2013}
{Stee} P.,  et~al., 2013, \mn@doi [Astronomy \& Astrophysics]
  {10.1051/0004-6361/201220302}, \href
  {https://ui.adsabs.harvard.edu/abs/2013A&A...550A..65S} {550, A65}

\bibitem[\protect\citeauthoryear{{Suffak}, {Jones}  \& {Carciofi}}{{Suffak}
  et~al.}{2022}]{M_Suffak_2022}
{Suffak} M.,  {Jones} C.~E.,   {Carciofi} A.~C.,  2022, \mn@doi [Monthly
  Notices of the Royal Astronomical Society] {10.1093/mnras/stab3024}, \href
  {https://ui.adsabs.harvard.edu/abs/2022MNRAS.509..931S} {509, 931}

\bibitem[\protect\citeauthoryear{{Toonen}, {Boekholt}  \& {Portegies
  Zwart}}{{Toonen} et~al.}{2022}]{S_Toonen_2022}
{Toonen} S.,  {Boekholt} T.~C.~N.,   {Portegies Zwart} S.,  2022, \mn@doi
  [Astronomy \& Astrophysics] {10.1051/0004-6361/202141991}, \href
  {https://ui.adsabs.harvard.edu/abs/2022A&A...661A..61T} {661, A61}

\bibitem[\protect\citeauthoryear{{Townsend}, {Owocki}  \& {Howarth}}{{Townsend}
  et~al.}{2004}]{R_H_D_Townsend_2004}
{Townsend} R.~H.~D.,  {Owocki} S.~P.,   {Howarth} I.~D.,  2004, \mn@doi
  [Monthly Notices of the Royal Astronomical Society]
  {10.1111/j.1365-2966.2004.07627.x}, \href
  {https://ui.adsabs.harvard.edu/abs/2004MNRAS.350..189T} {350, 189}

\bibitem[\protect\citeauthoryear{{Van Leeuwen}}{{Van
  Leeuwen}}{2007}]{F_vanLeeuwen_2007}
{Van Leeuwen} F.,  2007, \mn@doi [Astronomy \& Astrophysics]
  {10.1007/978-1-4020-6342-8}, \href
  {https://ui.adsabs.harvard.edu/abs/2007ASSL..350.....V} {350}

\bibitem[\protect\citeauthoryear{{Wheelwright}, {Bjorkman}, {Oudmaijer},
  {Carciofi}, {Bjorkman}  \& {Porter}}{{Wheelwright}
  et~al.}{2012}]{H_Wheelwright_2012}
{Wheelwright} H.~E.,  {Bjorkman} J.~E.,  {Oudmaijer} R.~D.,  {Carciofi} A.~C.,
  {Bjorkman} K.~S.,   {Porter} J.~M.,  2012, \mn@doi [Monthly Notices of the
  Royal Astronomical Society]
  {10.1111/j.1745-3933.2012.01241.x10.48550/arXiv.1202.4561}, \href
  {https://ui.adsabs.harvard.edu/abs/2012MNRAS.423L..11W} {423, L11}

\bibitem[\protect\citeauthoryear{{Wisniewski}, {Draper}, {Bjorkman}, {Meade},
  {Bjorkman}  \& {Kowalski}}{{Wisniewski} et~al.}{2010}]{J_P_Wisniewski_2010}
{Wisniewski} J.~P.,  {Draper} Z.~H.,  {Bjorkman} K.~S.,  {Meade} M.~R.,
  {Bjorkman} J.~E.,   {Kowalski} A.~F.,  2010, \mn@doi [The Astrophysical
  Journal] {10.1088/0004-637X/709/2/1306}, \href
  {https://ui.adsabs.harvard.edu/abs/2010ApJ...709.1306W} {709, 1306}

\bibitem[\protect\citeauthoryear{{Zorec} et~al.,}{{Zorec}
  et~al.}{2016}]{J_Zorec_2016}
{Zorec} J.,  et~al., 2016, \mn@doi [Astronomy \& Astrophysics]
  {10.1051/0004-6361/201628760}, \href
  {https://ui.adsabs.harvard.edu/abs/2016A&A...595A.132Z} {595, A132}

\bibitem[\protect\citeauthoryear{{Zorec} et~al.,}{{Zorec}
  et~al.}{2017}]{J_Zorec_2017}
{Zorec} J.,  et~al., 2017, \mn@doi [Astronomy \& Astrophysics]
  {10.1051/0004-6361/201628761}, \href
  {https://ui.adsabs.harvard.edu/abs/2017A&A...602A..83Z} {602, A83}

\makeatother
\end{thebibliography}

    \appendix

    \section {Derivation of the Theoretical Limits of the PMa} \label{sec:derivation}

        In order to obtain a theoretical, lower bound binary separation detectable via the PMa, it is easiest to consider a simple binary system. Take, for example, a system of two intermediate mass (i.e. $2 - 5\si{M_{\odot}}$) main sequence stars in a circular orbit. In such a system, where both stars are of a known mass, the centre of mass is easily derivable. Likewise, by applying the mass-luminosity relationship:
            
        \begin{equation}
            \label{eqn:mlr}
            \frac{L}{L_\odot} \simeq 1.4\frac{M}{M_\odot}^{3.5}
        \end{equation}
        
        for such stars, as given by \cite{S_C_Griffiths_1988}, the location of the photocentre is similarly able to be determined.
            
        The minimum theoretically detectable separation will occur for a system where (a) the difference between the centre of mass and the photocentre is at a maximum and, (b) where the change in absolute orbital phase is at a maximum (i.e the observations are spaced half an orbit apart).
        
        The latter of these criteria (b) is fairly easy to evaluate, provided we know the masses of the two stars. However, we must first evaluate the former criteria (a) in order to determine the mass ratio at which the largest offset between the photocentre and centre of mass is achieved. This proves to be a somewhat more involved process.
        
        By normalising the separation between the two stars to $1$, it is possible to write the photocentre and centre of mass equations as:
            
        \begin{equation}
            \label{eqn:norm_sep}
            a_{1, M/L} + a_{2, M/L} = 1
        \end{equation}
        
        \begin{equation}
            \label{eqn:com}
            M_1 a_{1, M} = M_2 a_{2, M}
        \end{equation}
            
        \begin{equation}
            \label{eqn:col}
            L_1 a_{1, L} = L_2 a_{2, L}
        \end{equation}
                
        \begin{equation}
            \label{eqn:com_col_offset}
            \delta = \lvert a_{1, M} - a_{1, L}\rvert
        \end{equation}
        
        where $a_{1/2, M/L}$ is the separation of the primary ($1$) or secondary ($2$) companion from the centre of mass ($M$) or photocentre ($L$) depending on the subscript, $M$ is the mass of an individual star, $L$ is the corresponding luminosity and $\delta$ is the offset between the centre of mass and the photocentre.
            
        From this set of equations, alongside the mass-luminosity relation given by Equation \ref{eqn:mlr}, it is possible to rewrite Equations \ref{eqn:com} and \ref{eqn:col} in terms of $a_{1, M/L}$ and the mass ratio of the two stars, $M_1 / M_2$, giving the expression of $\delta$ as:
            
        \begin{equation}\label{eqn:offset_mass_ratio}
            \delta = \lvert\frac{M_2}{M_1}(\frac{M_2}{M_1}^{3.5} + 1)^{-1} - \frac{M_2}{M_1}(\frac{M_2}{M_1} + 1)^{-1}\rvert
        \end{equation}
            
        By taking the derivative of this with respect to the mass ratio, and solving for $d\delta/d(\frac{M_1}{M_2}) = 0$, it is possible to obtain the mass ratio for which the maximum offset between the centre of mass and photocentre is achieved. This is found to occur for a mass ratio, $M_1 / M_2 = 0.47$ for intermediate mass main sequence stars.

        If we now assume a circular orbit with a stationary centre of mass, then the photocentre motion is resultant entirely from the orbital motion of the system. Thus, as the maximum change in proper motion occurs for observations separated by $\frac{3n + 1}{2}$ orbital periods, where $n$ is some integer greater or equal to zero, in this case resulting in the change in proper motion being twice the instantaneous proper motion. Thus, we can simply express this change in proper motion, $\Delta\mu$ as:
            
        \begin{equation}
            \label{eqn:tangential_vel}
            \Delta\mu = 2\mu_{st} = \omega r \delta  = \frac{2\pi r}{P}\delta 
        \end{equation}
            
        where $\mu_{st}$ is the short term proper motion measurement, which here can be reduced to the tangential velocity of the photocentre. $P$ is the period of the orbit and $r$ is the binary separation.

        While this approximation of the relation between the change in proper motion and the period of a binary system may seem to imply that the proper motion changes increasingly towards shorter period binaries, this approximation ignores the effects of smearing on the actual observationaly data. In reality this smearing across the observational timeframe causes a drop in sensitivity at shorter periods \citep{P_Kervella_2021}, with the maximum sensitivity being achieved for periods similar to twice that of the time elapsed between the two observations. Here, this resolves to a period of $\approx 50\si{yrs}$ for the Hipparcos-Gaia DR3 PMa, and $\approx 1\si{yr}$ in the case of the Gaia DR2-DR3 PMa.
            
        Using this calculated PMa, in concert with standard errors for the three surveys in question (in this case the Hipparcos re-reduction \citep{F_vanLeeuwen_2007} and the Gaia DR2 and 3 \citep{L_Lindegren_2018b, L_Lindegren_2021} although other values can be used as and where appropriate), a lower limit of separation can be obtained. This is achieved via propagating the standard errors through the calculations required to determine the PMa (as detailed in \cite{P_Kervella_2021}) and using this to generate a signal-to-noise ratio. As we have an expression for the change in proper motion that is a function of the separation we can thus write:

        \begin{equation}
            \frac{\Delta\mu}{N} =\frac{2\pi r}{NP}\delta
        \end{equation}

        where $N$ is the standard error of calculating the PMa. Thus by setting the above to equal the \cite{P_Kervella_2021} defined detection threshold of $3$, and solving for $r$ we arrive at a theoretical minimum binary separation detectable by the PMa. In the case of the PMa obtained from Hipparcos re-reduction and Gaia DR3 data this lower limit is estimated to be $0.01\si{''}$, and in the case of the Gaia DR2-DR3 PMa this is $0.0025\si{''}$. However the effect of the observational smearing is underestimated, particularly in the case of the Gaia DR2-DR3 PMa, where the PM measurement period covers the entirety of long term proper motion's timeframe - likely resulting in less sensitivity to small separations than predicted here.

    \bsp
    \label{lastpage}
\end{document}